# Deep Sequential Feature Learning in Clinical Image Classification of Infectious Keratitis

Yesheng Xu[a)*], Ming Kong[b)*], Wenjia Xie[a)*], Runping Duan[a)], Zhengqing Fang[b)], Yuxiao Lin[b)], Qiang Zhu[b)], Siliang Tang[b)], Fei Wu[b)], Yu-Feng Yao[a)]

[a] Department of Ophthalmology, Sir Run Run Shaw Hospital, Zhejiang University School of Medicine, Hangzhou, 310016
[b] College of Computer Science and Technology, Zhejiang University, Hangzhou, 310027
[*] Equal contribution

*Abstract*—Infectious keratitis is the most common entities of corneal diseases, in which pathogen grows in the cornea leading to inflammation and destruction of the corneal tissues. Infectious keratitis is a medical emergency, for which a rapid and accurate diagnosis is needed for speedy initiation of prompt and precise treatment to halt the disease progress and to limit the extent of corneal damage; otherwise it may develop sight-threatening and even eye-globe-threatening condition. In this paper, we propose a sequential-level deep learning model to effectively discriminate the distinction and subtlety of infectious corneal disease via the classification of clinical images. In this approach, we devise an appropriate mechanism to preserve the spatial structures of clinical images and disentangle the informative features for clinical image classification of infectious keratitis. In competition with 421 ophthalmologists, the performance of the proposed sequential-level deep model achieved 80.00% diagnostic accuracy, far better than the 49.27±11.5% diagnostic accuracy achieved by ophthalmologists over 120 test images.

## 1. Introduction

TRADITIONALLY, triage and diagnosis of diseases are carried out by physicians through observation based upon experience and knowledge constructed by individuals. For last couple of years, algorithms of deep learning through deep convolutional neural networks have been tested for medical imaging interpretation with significant advances. The application of algorithms for triage and diagnosis for diseases has been mainly tested in the fields which widely applied medical imaging technologies, including computerized tomography, magnetic resonance imaging, fundus photography, optical coherence tomography and pathologic images [1]. This is because medical imaging technology exported naturally rich image data, and commercialized medical imaging technologies create standardized and consistent medical images which can be collected in a short period of time in single institution or through multiple medical centers.

The diagnosis for many clinical diseases does not necessarily need commercialized medical imaging technologies, in which imaging recording is not routinely carried out in medical practice in many medical institutions, therefore collection of a large amount of image data will be dependent on historical accumulation sporadically dispersed in different medical centers. However, development of machine learning diagnostic systems for such diseases at least has equal importance. A study classifying skin lesion [2] offering malignant or benign judgement is a pioneer attempt in the field of not conventional sense of using medical imaging technologies. Corneal diseases may also be broadly classified in this category. Corneal diseases are a major cause of blindness worldwide [3] [4]. There are estimated 4.5 million individuals worldwide who suffer from moderate to severe vision impairment due to the loss of corneal clarity after corneal diseases [4]. Infectious keratitis is the most common cause to corneal diseases [5]. Normal cornea possesses unique characteristics of transparency. The uttermost feature of infectious keratitis is the pathogen growth in the cornea leading to focal mass cloudiness and the cornea roughness, ineluctably bringing out unique characteristics of each pathogenic microorganism for its growth in the tissue [6]. Diagnosis of an infectious keratitis mostly depends on discriminatively identifying the visual features of the infectious lesion in the cornea by ophthalmologists. Clinically, ophthalmologists routinely depend on slit lamp microscope to observe the normality or abnormality of the cornea and beyond. Besides as an observational tool, slit lamp microscope can also simultaneously be used to take a photograph and to record the existing status of the corneal manifestations for each patient visit, resulting in the development of a well annotated dataset for an artificial intelligence (AI) based infectious keratitis recognition and analysis.

Since 1998, we have developed a large, well annotated slit lamp microscopic image dataset of 115, 408 images in total from 10, 609 corneal disease patients. The collected dataset enabled us to devise one deep learning based method to perform infectious keratitis diagnosis in an end-to-end manner. In order to intuitively mimic the way how ophthalmologists diagnose infectious keratitis, we proposed a feature learning mechanism to identify the informative visual patterns via a manner of the *sequential-level* feature learning, which means the sampled patches from the center to the edge of the infectious lesion area in the clinical picture are grouped into a sequence of ordered set and fed into neural network for feature learning. We argue that the proposed sequential-level feature learning mechanism can utilize the spatial relationship among patches from the infectious lesion area and is able to disentangle exploratory factors of variations underlying the data sample and is a potential strategy to achieve a more reliable and effective performance for accurate diagnosis (in Figure 1).

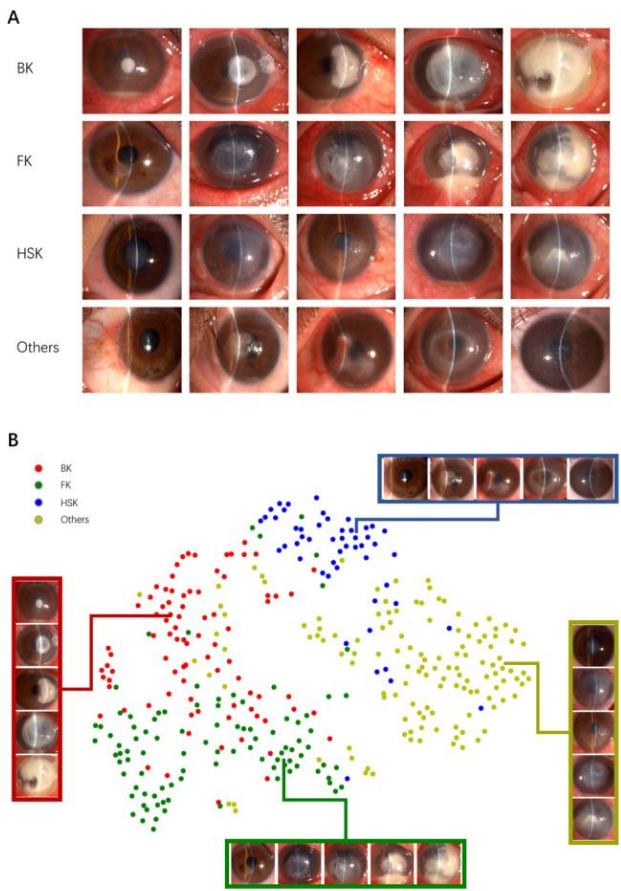

Fig. 1. The representative slit-lamp microscopic images and the representations of t-SNE visualization of the embedding features in the proposed Sequential-Ordered Sets (SoS) model for the four classes of the corneal diseases. A is the representative slit-lamp microscopic images of bacterial keratitis (BK), fungal keratitis (FK), herpes simplex viral stromal keratitis (HSK), and the others including those except the above three mentioned categories of corneal diseases. They exhibit different visual features at different stages of the same disease or show difference of visual features among categories. B shows the deep features learned by the proposed SoS model being embedded into a two-dimensional space via t-SNE for each category of the disease. t-SNE is utilized for visualizing the high-dimensional data which is the feature representation in the SoS model of the diagnosis-proven photographic test sets (362 images). Colored point clouds represent the different categories of the diseases, showing how the algorithm groups the diseases into different clusters. Insets show images corresponding to various points.

Our model was evaluated using the dataset and achieved an accuracy of correct diagnosis beating more than 400 ophthalmologists.

## 2。RELATED WORKS

### A. Medical Data Mining

Over the years, Electronic Medical Records (EMR) have accumulated large quantities of medical data, which enables researchers to discover the underlying knowledge. Data mining methods have been widely used on medical data to discover hidden knowledge and to use the extracted knowledge to help the prediction, diagnosis and treatment of various harmful diseases.

Disease predicting is significant in preventing the disease from happening and reducing the harm. Yang et al. uses patients' health records to forecast the potential diabetes complications as well as to discover the underlying association between complications and lab test types [39]. He et al. predicts lung cancer postoperative complication using EMR dataset and extracts crucial variables from the dataset simultaneously [40].

EMR with predicted diagnostic labels and medication information can help automatic assistant to predict disease diagnosis and provide a rapid diagnostic reference for doctors. Ni et al. uses a large EMR text dataset to model the context of EMR of each disease and performs an accurate disease diagnosis prediction in EMR [41]. Wright et al. uses data mining methods to get useful relations and rule sets from the medical datasets to predict which medication is prescribed next [38].

### B. Traditional Shallow Models in Medical Image Application

Traditional method used hand-craft features (in general shallow models) for medical image classification and segmentation. Scott et al. in 2003, used gradient orientation, corner and edge strength for the detection of vertebrae in dual energy X-ray images [42]. Region splitting and merging technique is well known in region-based approach. This technique is a combination of region splitting and merging. Manousakes applied splitting and merging technique in trying to overcome the difficulties occurred when using homogeneity measures on MRI [43]. Y.Q Zhang et al. introduced basic mathematical morphological theory and operations, the novel mathematical morphological edge detection is proposed to detect the edge of lungs CT image with salt-and-pepper noise [44]. The experimental result shows that the method proposed is more efficient for both medical image de-noising and edge detection than the best edge detection method in 2006. Kaus et al. use K-means Cluster to automatically do segmentation of the left ventricle in cardiac MRI [45]. D. Cordes et al. had done a research by using hierarchical clustering to measure connectivity in functional MRI [46]. This method is able to detect similarities of low-frequency fluctuations, and the results indicated that the patterns of functional connectivity can be obtained with hierarchical clustering that resembles known neuronal connections. In 2006, Pohl et al. presented a method of embedding the signed distance maps into the linear Log Odds space, which could solve the modeling problems [47]. Although these methods focused on region, edge and clustering, they have limited performance on real-world data [48]

### C. Deep-Learning Methods in Medical Image Application

In the development of computer aided diagnosis, deep learning is now widely used in medical image recognition [36,37]. The basic structure of deep learning is convolutional neural network (CNN), which has three types of the layer including convolution, pooling, and total connection. To develop a robust AI algorithm based on CNN, we usually need a large amount of annotated data.

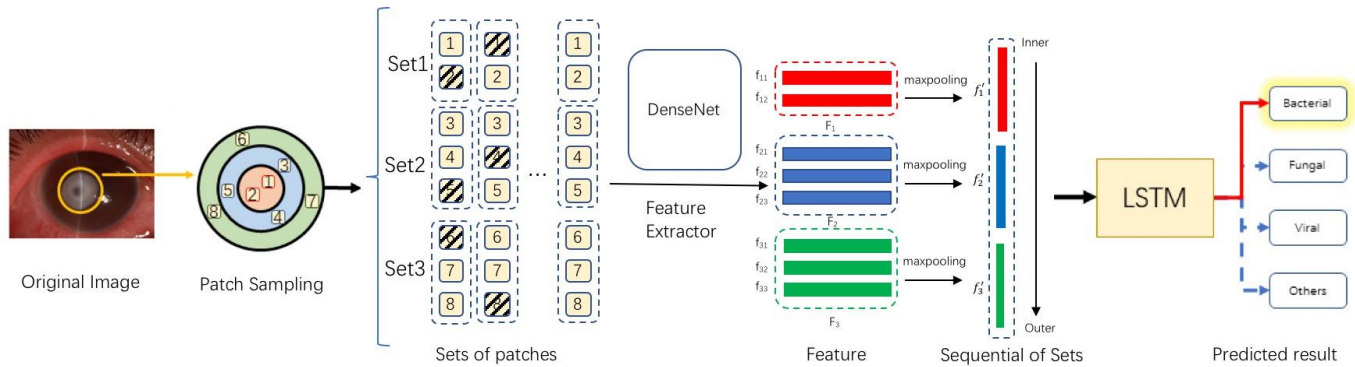

Fig. 2. The process of sequential deep feature learning for one lesion area. For each slit-lamp microscopic image, lesion area is divided into the minimum circumscribed circle to K circular ring parts (k=3 here only for intuitive clarification). From the inner to the outer of circular rings, we sample patches from each circular ring, and the sampled patches are used to generate sequence of sets. The sequential features can be learned via max-pooling and LSTM.

However, the standardized collection of medical images is not as easily as collecting general natural images. Nowadays, several public medical image databases and multicenter collections of data can help to solve the problem. Some types of medical image data such as X-rays, computerized tomography (CT), electrocardiograph, and pathology images can be collected in a large quantity. By using these big data, the recent CNN-based AI algorithms can perform anatomical structure segmentation on CT images [24], classify normal or abnormal findings of chest radiographs [25], perform screening for lung or breast cancer [26,27], detect critical findings in head CT scans [28], classify liver lesion using GAN based model [29], perform screening for heart conditions [30,31], and detect lymph node metastases in pathology images [31,32].

In the field of ophthalmology, due to the easily collecting of images from fundus photography and optical coherence tomography (OCT), the major area in which CNN-based AI algorithms have been applied was detecting retinal diseases such as diabetic retinopathy, age-related macular degeneration and glaucoma [33-35].

At present, AI-assisted medical diagnostic systems are mainly applied in the field of aforementioned medical imaging. For the diagnosis of disease that relies on the use of natural observation, it mainly depends on the personal experience of the doctor. One example is skin lesion, the current AI algorithm can differentiate malignant melanoma from benign lesions on digital skin photographs [2]. Corneal disease is another example, that ophthalmologists may use of a slit-lamp microscope to achieve the right diagnosis. So far, there is no research that has introduced AI technology to improve the diagnostic accuracy of corneal disease.

## II. MEHTODS

*A. Image Datasets*

Upon an institutional review board approval, the image dataset for this study included 115,408 clinical digital images taken from 10,609 patients with 89 categories of corneal diseases by slit lamp microscopies during the time of period from May of 1998 to present in the Department of Ophthalmology, Sir Run Run Shaw Hospital, Zhejiang University School of Medicine. The clinical images were taken by two types of slit lamp microscopes, i.e., ZEISS slit lamp microscope SL 130 (Carl Zeiss Meditec AG., Jena, Germany) integrated with the SL Cam for Imaging Module providing each image with a resolution of 1024 x 768 pixels and Topcon slit lamp microscope (TOPCON Corp., Tokyo, Japan) affiliated with Digital Camera Unit DC-1 offering an image resolution of 1740 x 1536 pixels or 2048 x 1536 pixels, respectively.

In the dataset, images taken from patients with corneal infection at the active stage, including bacterial keratitis, fungal keratitis and herpes simplex viral stromal keratitis(HSK), were selected for the training or testing set for the algorithmic classification into each infectious category. All the images from the patients with corneal infections were annotated with a definite clinical diagnosis that was corroborated by at least two pieces of the following evidences: 1) the clinical manifestations of the corneal infection as shown in the photos of Fig. 1A, 2) the progression of the corneal infection was influenced and terminated by diagnostic pertinence single drug or combined drugs therapy leading to its ultimate curation, 3) pathogen identification of the sample from the infection site: in bacterial and fungal infections, pathogenic diagnosis either confirmed by sample smear under microscopic examination or organism culture, and in viral infection, pathogenic diagnosis confirmed by PCR evaluation of samples from the tear or corneal scraping tissues. In addition to categories of the corneal infections, images taken from patients suffering from other corneal diseases with similar visual features were classified into the category of other diagnosis. This category includes varieties of corneal dystrophies, phlyctenular keratoconjunctivitis, various corneal tumors, corneal papilloma, corneal degeneration, and even acanthamoeba keratitis. Representative image series in each category are shown in Fig. 1A.

The final data-set involved 2,284 images from 867 patients for this study. The training set consisted of randomly selected 387 images of bacterial keratitis, 519 images of fungal keratitis, 488 images of HSV stromal keratitis, and 528 images of other corneal diseases, from 747 patients. The testing set consisted of randomly selected 86 images of bacterial keratitis, 97 images of fungal keratitis, 51 images of HSV stromal keratitis, and 128 images of other diagnosis, from 120 patients. In order to evaluate ophthalmologists' classification performance, the first-time diagnosis images of each patient in the testing set were selected to build up a dataset for evaluation of ophthalmologists (i.e., 120 images in total had been used to evaluate the performance of ophthalmologists).

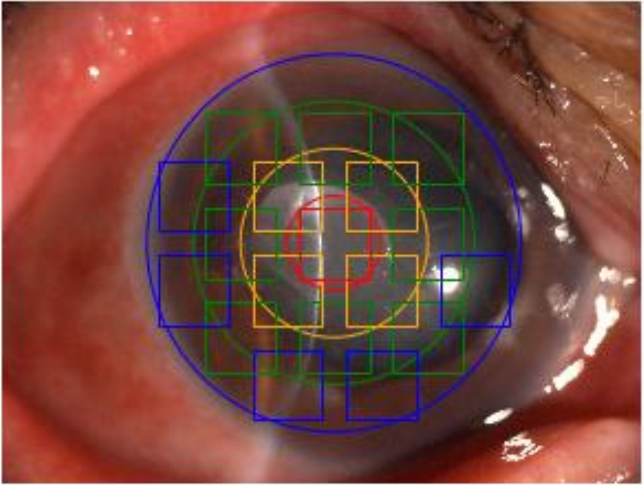

Fig. 3. An illustration of how patches were sampled and how they were divided into K sets. Circles represent the boundaries for each set and squares represent the sampled regions. Notice that to avoid too much overlapping on the picture, only half of the patches are shown in this picture.

### B. Sequential-level feature learning based diagnostic deep models

As aforementioned, we devised a sequential-level feature learning method for the classification of infectious keratitis. In order to demonstrate the superiority of the proposed method, we compare our proposed method with other different models, namely the *image-level* feature learning and the *patch-level* feature learning.

The image-level feature learning deep model uses a transfer learning technique to solve the problem of limited training data [7], [8], in which the whole clinical images without annotation are applied directly to a CNN for diagnostic analysis and classification. In our experiments, we chose three classic architectures for image classification: VGG16 [9], GoogLeNet-v3 [10], and DenseNet [11].

In the patch-level feature leaning deep model, the image of the anterior segment of an eye is initially annotated by manual drawing dividing the image into four parts: the infectious lesion area of the cornea, the area beyond the lesion of the cornea, the injection of conjunctiva, and the exudation of the anterior chamber, respectively. There are three transfer learning architectures in this deep model, i.e., VGG16, GoogLeNet-v3 and DenseNet. After each patch is classified, a method of majority voting is implemented to predict the classification result of each clinical image.

In the sequence-level feature learning model, for each image the focus of attention is placed on the lesion, if there is any. The centroid of the lesion is annotated for building a minimum circumscribed area. The minimum circumscribed area is further divided into K circular rings scaling up around the center. The partitioning method is illustrated in Fig. 2. From the inner to the outer circular rings, the sampled patches inside the i-th circular ring is used to build up a set denoted patches $S_i$ and a sequence of sets $\{S_1, S_2, ..., S_K\}$ following the order from the innermost to the outermost. To address the issue of limited annotated data, in the training process, a drop-out mechanism of randomly dropping out elements from each set is applied, which can generate more sequences of the sets, enabling expanding the data diversity and making the trained model more robust.

Each patch in a set is applied to a deep residual CNN (i.e, DenseNet) through a sequential feature learning manner via encoder-decoder framework [12]-[14]. The convolutional structured encoder can transform the j-th patch in the i-th set $p_{ij}$ into a vectorial feature $f_{ij}$ to describe its innate characteristics, represented as a set of patch-level features $\{F_1, F_2, ..., F_K\}$. For each set $F_i$, a combined overall patch features through operating a Max Pooling calculation can be generated represented as the set $f'_i$, which represents the global characteristic over a given set. Since the sets from the innermost to the outermost of a lesion consist of a sequence of sets, a long short-term memory (LSTM) [12], one of the most classic models to learn sequential data in deep learning, can be used to transfer the set feature sequence $\{f'_1, f'_2, ..., f'_K\}$ into a representation for the classification. The features for the images can be decoded by a fully-connected network layer, and the probability of each category of corneal diseases is described by a Softmax calculation for the learned features. Fig. 1B illustrates the embedding features of each lesion in a 2-dimensional space. The working system is shown in Fig. 2. Comparing the result of the predicted probability with the ground-truth type of keratitis, the loss from the result is back-propagated to fine-tune parameters of the model [15], [16].

### C. Recruitment of Ophthalmologists for Image Based Diagnostic Analysis

Ophthalmologists had been recruited all over the country of China to test their performances for image based diagnostic analysis as a comparison study with the developed deep learning methodology. The images presented to the recruited ophthalmologists were randomly selected from the first visit and diagnosis-proven images of each patient in the testing set (i.e., in total 120 images). The recruited ophthalmologists varied in different academic titles (from residential ophthalmologists to senior ophthalmologists all the way to full clinical professors in medical schools), in different affiliations (from teaching hospitals in university medical schools to public municipal hospitals to community clinics), and in different professional experiences (categorized into 1-5 years, 6-10 years, 11-15 years, 16-20 years, and over 20 years). In total, we had recruited 421 ophthalmologists all over the country in China.

The ophthalmologist manual examination for image based diagnostic analysis followed the two-step protocol. In the first step, an ophthalmologist conducted an image only diagnosis, where images of four categories of corneal diseases from the first-time

TABLE I
THE PERFORMANCE OF CLASSIFICATION ACCURACY AMONG DIFFERENT DEEP LEARNING MODELS ON TEST DATASET (362 IMAGES IN TOTAL, DATA PRESENTED IN PERCENTAGE)

|  | Algorithm | Test dataset | | | | |
| --- | --- | --- | --- | --- | --- | --- |
|  |  | Acc | BK | FK | HSK | Others |
| Image Level | VGG16 (Image) | 55.24 | 48.84 | 52.57 | 62.74 | 53.90 |
|  | GoogLeNet-v3 (Image) | 57.73 | 53.49 | 55.67 | 66.67 | 58.59 |
|  | DenseNet (Image) | 61.04 | 60.46 | 56.70 | 80.39 | 57.03 |
| Patch Level | VGG16 (Voting) | 52.50 | 45.34 | 54.64 | 56.00 | 54.68 |
|  | GoogLeNet-v3 (Voting) | 55.52 | 44.19 | 51.55 | 74.51 | 58.59 |
|  | DenseNet (Voting) | 66.30 | 59.30 | 68.04 | 58.82 | 72.66 |
| Sequence Level | Random-Ordered Patches (ROP) | 74.23 | 75.29 | 68.04 | 82.35 | 75.00 |
|  | Sequential-Ordered Patches (SOP) | 75.14 | 66.28 | 86.60 | 84.31 | 68.75 |
|  | Sequential-Ordered Sets (SOS) | 78.73 | 65.12 | 83.51 | 90.20 | 79.70 |

The image-level feature, the patch-level feature and the sequential feature are learned from the whole image, or from the lesion area, or from the sequence of patch set, respectively. The test dataset consists of 362 images including 86 bacterial keratitis, 97 fungal keratitis, 51 HSV stromal keratitis and 128 others from 120 • patients. Acc shows the overall accuracy of each model, and column BK, FK, HSK and Others show the recall on each category respectively.

diagnosis images of each patient in the testing set, specifically bacterial keratitis, fungal keratitis, HSV stromal keratitis, and other corneal diseases, were presented to the ophthalmologist, and for each image, a diagnostic decision was made by the ophthalmologist through the manual examination. Then in the second step, the ophthalmologist was further provided with additional standardized and structured medical information affiliated to each image including brief medical history, time of onset, the grade of pain and recurrence episodes if any, and history of drug use. The ophthalmologist was then asked to make a diagnostic decision for each image through the manual examination of the image together with the additional medical information. All ophthalmologists conducted the manual examination to reach the diagnostic decision for each testing image independently without time limitation.

*D. Statistical Analysis*

Given the different confidences resulting from the different academic titles, different affiliations, and different professional experiences, the Statistical Package for Social Sciences (SPSS Version 18.0; Cary, NC, USA) was used for statistical analysis for the ophthalmologist manual diagnosis data. The average performances as the diagnosis accuracy achieved by the ophthalmologists were summarized and represented in terms of mean ± standard deviation in percentage. Data normality was initially verified using the Kolmogorov-Smirnov test. Differences in the diagnosis accuracy among different hospital levels and professional title groups were analyzed using one-way analysis of variance (ANOVA), in accordance with the data normality. Least significant difference (LSD) was used for post-hoc analysis of the parametric variables. Correlation between the diagnosis accuracy and the years of professional experiences was tested using the Pearson's correlation coefficient. Multi-linear regression analysis with the stepwise method was employed to explore the influence of the demographic factors in terms of academic titles, hospital levels, and years of professional experiences. Paired t-test (for normally distributed variables) or Wilcoxon signed ranks test (for non-normally distributed variables) were performed to determine if there were any significant differences in diagnostic accuracy between the doctors' performances with and without additional medical information. The significance level for all the tests is set to 0.05.

III. RESULTS

*A. Performances of the different Deep Models*

Image-level deep model is currently popular for clinical image diagnosis, in which the whole clinical images are directly applied to CNNs. Three classic deep architectures, VGG16, GoogLeNet-v3 and DenseNet, are used, respectively, in this model to report the diagnostic performances of this model for bacterial, fungal, and HSV stromal keratitis documented in Table I. Considering the fact that the whole image directly applied to CNN in the training process may contain irrelevant information, we thereafter developed the patch-level deep model [17], [18] using architectures of VGG16, GoogLeNet-v3, and DenseNet, respectively. In the patch-level deep model, instead of using whole image, patches including infectious lesion of the cornea, beyond infectious lesion of the cornea, the injection of conjunctiva, and the exudation of the anterior chamber are initially annotated by manually segmentation. We have found that the three patch-level models can secure an accuracy of 49.62%, 51.52%, and 60.00%, respectively, for patch classifications (i.e., the classification of each patch into a corresponding infectious keratitis). After each patch is classified, a majority voting is implemented to perform clinical image classification. The patch-level deep models with voting have respectively achieved the accuracy of 52.50%, 55.52%, and 66.30% as documented in Table I.

Finally, we applied the sequential-level deep models which are considered to have the ability of preserving the subtle spatial structures of clinical images. As aforementioned, the sequential-level features are learned in an inner-outer sequential order (referred to as the sequential-ordered set, SOS) and we have achieved 78.73% classification accuracy with SOS features. Instead of generating a sequence of sets in an inner-outer sequential order, we can also generate a sequence in terms of random-ordered patches (ROP) and sequential-ordered patches (SOP). ROP generates a sequence of patches via a random order and SOP generates a sequence of patches via an inner-outer order (but without the utilization of a set structure to group each patch into different sets). The final evaluation shows that ROP features have arrived at an accuracy of 74.23%（75.29% for bacterial keratitis, 68.04% for

TABLE II
DEEP LEARNING MODELS COMPETING WITH OPHTHALMOLOGISTS USING DATASET OF 120 IMAGES IN TOTAL (DATA PRESENTED IN PERCENTAGE)

|  | Algorithm | Dataset for Evaluation of Ophthalmologists | | | | |
|---|---|---|---|---|---|---|
|  |  | Acc | BK | FK | HSK | Others |
| Image Level | VGG16 (Image) | 50.83 | 46.67 | 43.33 | 73.33 | 40.00 |
|  | GoogLeNet-v3 (Image) | 55.83 | 50.00 | 63.33 | 70.00 | 40.00 |
|  | DenseNet (Image) | 64.17 | 56.67 | 63.33 | 80.00 | 56.67 |
| Patch Level | VGG16 (Voting) | 51.67 | 23.33 | 43.33 | 76.67 | 63.33 |
|  | GoogLeNet-v3 (Voting) | 54.17 | 26.67 | 73.33 | 80.00 | 36.67 |
|  | DenseNet (Voting) | 71.67 | 46.67 | 86.67 | 73.33 | 80.00 |
| Sequence Level | Random-Ordered Patches (ROP) | 77.50 | 66.67 | 70.00 | 93.33 | 80.00 |
|  | Sequential-Ordered Patches (SOP) | 79.17 | 73.33 | 70.00 | 96.67 | 76.67 |
|  | Sequential-Ordered Sets (SOS) | 80.00 | 53.33 | 83.33 | 93.33 | 90.00 |
| Human Level | Average Performance of Ophthalmologists provided with image only | 49.27 | 46.55 | 45.56 | 65.01 | 39.95 |
|  | Average Performance of Ophthalmologists provided with image together with medical history | 57.16* | 55.55* | 56.28* | 73.25* | 43.56* |

The first-visit and diagnosis-proven images of patients with 4 categories of the corneal diseases were selected from the testing set to build up a dataset for evaluation and competition of deep learning models with ophthalmologists. The dataset included 120 clinical images. * P < 0.001 compared to the average performance of ophthalmologists provided with image only.

fungal keratitis and 82.35% for HSV stromal keratitis）, and SOP features have arrived at an accuracy of 75.14%）. This evaluation study has demonstrated that the sequential-level deep models are the best models for automatic, image only diagnosis for corneal diseases.

*B. Comparison Study on Ophthalmologists' Diagnosis*

We evaluated all the algorithms that we considered in this paper via the dataset to compare the performances between each algorithm and ophthalmologists. Table II observes the performances of accuracy of all the algorithms and the average performance of ophthalmologists via this dataset (120 images). The performance of ophthalmologists in diagnosis of clinical images is summarized in Table III. There were 421 ophthalmologists recruited from all over China participating in this comparison study for the corneal diseases. The average performance of accuracy of total ophthalmologists without additional medical information was 49.27±11.5% (range: 20.00 to 86.67%), which was far more below than the accuracy performed by AI deep learning models, such as in the algorithm of Sequential-Ordered Set, the total diagnostic accuracy was 80.00%, including the accuracy of 53.33% for bacterial keratitis, 83.33% for fungal keratitis, and 93.33% for HSV stromal keratitis, respectively (Table II). Fig. 3 shows the receiver operating characteristic (ROC) curve and the confusion matrix of the Sequential of Sets (SoS) Model and performance of ophthalmologists. ROC curve is a visualization method for classification model. The area under curve (AUC) is the measure of performance, with a maximum value of 1. The model achieves superior performance against an ophthalmologist if the sensitivity-specificity point of the ophthalmologist lies below the curve of the classification model.

The working place effecting the ophthalmologists' performance was demonstrated in this study that those from teaching hospitals demonstrated a far better performance than those from city hospitals and community clinics (both P < 0.001), whereas no significant difference was found between city hospitals and community clinics (P = 0.226). The ophthalmologists with higher professional ranks appear to have a better performance in diagnosing clinical images with a better accuracy, such as the attending ophthalmologists and fellows were better than residents (P < 0.001 and P = 0.003, respectively), but no significant difference was found between the groups of attending from fellow ophthalmologists (P = 0.071). No significant correlation was found between the duration of employment and diagnostic accuracy (R = 0.084, P = 0.084).

TABLE III
THE PERFORMANCE OF CLASSIFICATION ACCURACY IN AVERAGE AND THOSE RELATED TO THE HOSPITAL LEVEL, YEARS OF EMPLOYMENTS AND PROFESSIONAL TITLES OF THE OPHTHALMOLOGISTS

| Dr. Group | Participant Num | Boxplot | Mean±STD (%) | Range (%) |
|---|---|---|---|---|
| **Total** | 421 | | 49.27±11.85 | [20.00, 86.67] |
| **Hosp. RK.** | | | | |
| Teaching | 84 | | 55.69±12.19 | [33.33, 86.67] |
| City | 171 | | 48.46±10.70 | [24.17, 78.33] |
| Community | 166 | | 46.84±11.63 | [20.00, 81.67] |
| **Year of Employments** | | | | |
| 1-5 | 89 | | 45.96±13.22 | [22.50, 78.33] |
| 6-10 | 117 | | 49.39±11.80 | [20.83, 76.67] |
| 11-15 | 69 | | 50.85±12.03 | [24.17, 81.67] |
| 16-20 | 41 | | 49.94±12.30 | [20.00, 76.67] |
| >20 | 105 | | 50.64±9.66 | [25.00, 86.67] |
| **Physician RK.** | | | | |
| Attending | 173 | | 51.76±11.94 | [20.00, 86.67] |
| Fellow | 150 | | 49.38±11.18 | [20.83, 81.67] |
| Resident | 98 | | 44.69±11.66 | [22.50, 78.33] |
| **Hosp. RK. & Physician RK.** | | | | |
| Attending in Teaching | 36 | | 57.08±12.02 | [33.33, 86.67] |
| Fellow in Teaching | 30 | | 55.47±12.21 | [35.00, 77.50] |
| Resident in Teaching | 18 | | 53.29±12.21 | [33.33, 75.00] |
| Attending in City | 78 | | 51.63±10.45 | [24.17, 77.50] |
| Fellow in City | 59 | | 46.96±9.07 | [24.17, 66.67] |
| Resident in City | 34 | | 43.80±11.57 | [25.00, 78.33] |
| Attending in Community | 59 | | 48.69±10.36 | [20.00 74.17] |
| Fellow in Community | 61 | | 48.72±12.53 | [20.83, 81.67] |
| Resident in Community | 46 | | 41.99±10.51 | [22.50, 63.33] |

(RK=Rank, * P = 0.003, ** P < 0.001)

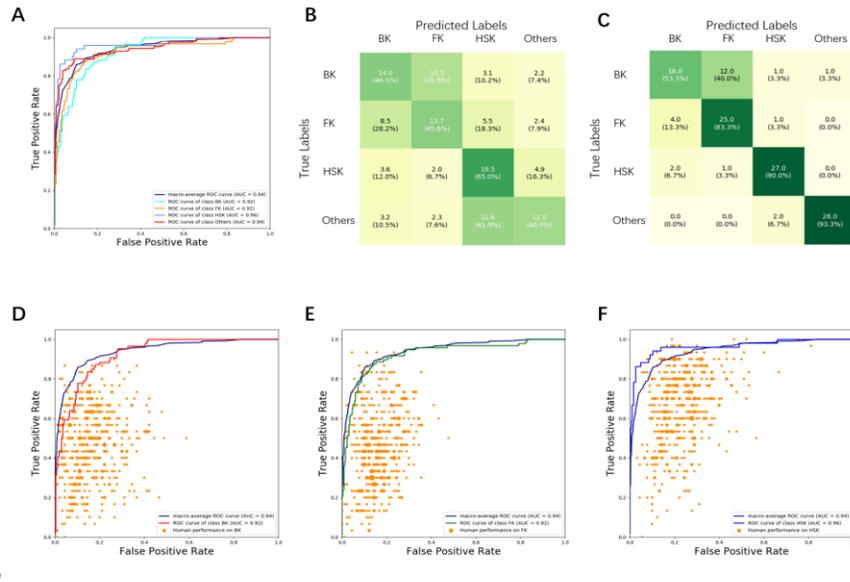

Fig. 4. The ROC curve and the confusion matrix of the Sequential of Sets (SoS) Model and performance of ophthalmologists. A is the receiver operating characteristic (ROC) curve of SoS model. B and C are the confusion matrix of ophthalmologists and SoS model on the Dataset for Evaluation of Ophthalmologists. D, E and F are the ROC curves for the disease category of bacterial keratitis (BK), fungal keratitis (FK) and HSV stromal keratitis (HSK).

When the factors of hospital ranking and doctor's ranks are considered together, better performance was found in the group of ophthalmologists with attending title from teaching hospitals (accuracy of 57.08±12.02%, range: 33.33 to 86.67%), than the poorer performance in the group of ophthalmologists with resident ranks from community clinics (accuracy of 41.99±10.51%, range: 22.50 to 63.33%). The stepwise multiple regression analysis resulted in three models that affected diagnostic accuracy. Model 1 ($R^2 = 0.062$) had the only factor of hospital levels ($\beta = 0.254$, $P < 0.001$); Model 2 ($R^2 = 0.100$) had the factors of hospital levels ($\beta = 0.239$, $P < 0.001$) and professional titles ($\beta = 0.200$, $P < 0.001$); Model 3 ($R^2 = 0.109$) had all three factors of hospital levels ($\beta = 0.227$, $P < 0.001$), professional titles ($\beta = 0.326$, $P < 0.001$) and years of employment ($\beta = -0.164$, $P = 0.024$).

When the ophthalmologists were further provided with additional medical information affiliated to each image including brief medical history, time of onset, the grade of pain and recurrence episodes if any, and history of drug use, the mean total diagnostic accuracy was raised from 49.27% to 57.16%, the difference was statistically significant (Wilcoxon signed ranks test, $P < 0.001$). In detail, the accuracy increased from 46.55% to 55.55% ($P < 0.001$) for bacterial keratitis, from 45.56% to 56.28% ($P < 0.001$) for fungal keratitis, and from 65.01% to 73.25% ($P < 0.001$) for HSV stromal keratitis, respectively. With additional medical information, the mean total accuracy of 404 doctors increased by 8.28%, 9 doctors decreased by 2.13%, and the accuracy of other 8 doctors remained unchanged.

## IV. Discussion

In general, fact judgement by human is achieved through perception of vision, audio, touch, taste and smell, which enables a person to make things classified into reasonably category [19]. Visual perception plays the most important role over others [20]. Physicians making diagnosis for diseases mainly depend on observation and reasoning. Among all human diseases, the corneal diseases have the most direct and the most sufficient displays of changes in visual perception, because the healthy cornea of an eye has exceptional characteristics of complete transparency, which is in sharp contrast to the pathological conditions that always present image changes in and beyond the cornea. Diagnostic decision making of corneal disease by human professionals is carried out through image understanding and analysis, which is probably the most appropriate task assumed by AI as an alternative or assistance to human.

Generally speaking, deep learning is driven by a large amount of annotated data [21], [22]. However, it is not clear how large the training data amount of clinical images should be that meets the requirement of developing an AI system for diagnosing clinical diseases. Our center has been collecting and documenting corneal disease cases with clinical images for 20 years. But when all the images are annotated according to each disease category, there can be thousands of clinical images in the most common disease categories, whereas there are only a few dozen of clinical images in some rare disease categories. The imbalance of the annotated data in each corneal disease category leads us to have to focus on the most common diseases, such as infectious corneal diseases to develop a first stage of AI diagnostic system in this study.

In this study, we have demonstrated that deep learning through convolutional neural networks can be applied to clinical diagnosis for corneal infectious diseases using clinical images taken by slit lamp microscopy. We have evaluated three sets of nine deep learning architectures in total as an effort to develop an image only diagnostic system for corneal infectious diseases. From the results of image-level and patch-level models, we can say that despite only having 4 categories, this is a hard problem, especially for VGG16 and GoogLeNet-v3. These two structures had poor performance in patch classification, and as a result, voting among patches did not improve their performance much. On the contrary, DenseNet reached 60.00% in patch classification, and achieved 66.30% after voting. This shows that focusing on patches from the infectious lesion area can result in higher performance than seeing the whole picture, provided that the model performs good enough in patch classification. The Random-Ordered Patches method can be viewed as another way to combine patch features besides voting. Its result shows that even without spatial information, patch-level model can be further improved given appropriate combining method. We have found that in overall the Sequentially-Ordered Set is the most promising method for image only diagnosis for corneal disease images. The possible reason for why Sequentially-Ordered Set was better than the other methods can be explained that an appropriate utilization of spatial structures of clinical images is directly implemented into this model of deep learning. Sequentially-Ordered Patches performed not as good because it did not take into account the circular structure of the lesion area. To our best knowledge, this is the first study that shows a deep learning model for corneal disease classification with higher accuracy than that of human ophthalmologists in image only diagnosis. It was noted in this study that in general professional human performance in image-only diagnosing corneal diseases was worse than that of an AI system. There is no doubt that the incorrect diagnosis can lead to prolonged use of inappropriate medications that causes the identifying features being obscured[6], making human decision-making for diagnosis more difficult. The multiple regression analysis in our study demonstrated that the three demographic factors in terms of academic ranks, affiliations, and professional service duration had influences on the diagnostic performance, whereas the coefficients of determination were low in the three models. This indicates that the above factors may not truthfully and comprehensively determine the diagnostic accuracy of corneal diseases in ophthalmologists, or the factors affecting the diagnostic performance may be very complicated and may not be well and truly summarized simply by the above three factors. Therefore, if AI can assist clinicians improve their ability significantly with a higher diagnostic accuracy, this shall greatly benefit patients suffering from corneal diseases, saving medical resources, and reducing societal burden. There is still a large population of 4.5 million individuals who are now suffering from moderate to severe vision impairment due to the loss of corneal clarity after corneal diseases worldwide [4], especially in the developing countries. There are two ways that can raise diagnostic accuracy. One is to improve the physician training system and to strengthen the professional education and training for physicians; the other is to develop a practicable artificial intelligence system to assist diagnosis. Our current study demonstrates that it is realistically achievable to develop an AI system by using clinical images to improve the diagnostic accuracy for corneal diseases. In examining the ophthalmologists' performance, we found that, when the medical professionals were provided with images together with additionally medical history, the diagnostic accuracy was raised to a certain extent (from 49.27% to 57.16%, $P < 0.001$) as compared with that when the professionals were provided by images only for recognizing to make decision. This result indicates that, for medical professionals, recognizing image features is the most important key to improve their diagnostic accuracy, however, for improving our AI diagnostic system to raise the diagnostic accuracy, a multi-modal learning model (i.e., the effective combination of visual and non-visual information) or a more suitable sequential learning model may necessarily be devised for the future work.

It is undeniable that our AI diagnosis accuracy at this stage is only confirmed by the limited image data we have collected, with the comparison study with the ophthalmologists' diagnosis performance using the same clinical images. A real-world application of such an AI system in assisting physicians in clinical practice requires further clinical evaluations in a larger scale extensively [23].

SUPPLEMENT

Application to the availability of the medical training/validation data that support the findings of this study may be available from the authors upon reasonable request and with permission of the Sir Run Run Shaw Hospital, whereas an appropriate anonymization is needed to ensure that the patients' confidentiality is absolutely protected. The evaluation system developed in this study using diagnosis-proven first visit clinical image data and medical record data for evaluating the ophthalmologists' performance and competition between ophthalmologists and AI diagnostic system can also be available from authors upon reasonable request.

ACKNOWLEDGMENT

We thank Zhongfei Zhang for discussions and comments.